\documentclass[final,5p,twocolumn,times,lefttitle]{elsarticle}
\biboptions{sort&compress}
\usepackage[hidelinks,bookmarksopen,bookmarksnumbered]{hyperref}
\makeatletter
\def\ps@pprintTitle{%
	\let\@oddhead\@empty
	\let\@evenhead\@empty
	\def\@oddfoot{}%
	\let\@evenfoot\@oddfoot}
\makeatother

\usepackage{fancyhdr}
\usepackage{pgfbaseimage}
\usepackage{orcidlink}
\pgfdeclareimage[height=0.75cm]{cc-license}{by-nc-nd}
\fancypagestyle{pprintTitle}{%
	
	\fancyhf{}
	\lfoot{\href{}{Published in the proceedings of \emph{Symposium Elektromagnetismus, Künzelsau, 2025}.\\TAE Tagungshandbuch 2025, \texttt{ISBN 978-3-943563-55-9}, pp. 53--58.} \\ © 2025. This manuscript is made available under a \href{https://creativecommons.org/licenses/by-nc-nd/4.0/}{CC-BY-NC-ND 4.0 license}.}%
	\rfoot{\vspace{3mm}\href{https://creativecommons.org/licenses/by-nc-nd/4.0/}{\pgfuseimage{cc-license}}}
	\lhead{}
}

\usepackage{cmap} 
\usepackage[T1]{fontenc}
\usepackage{textcomp}
\usepackage{cuted}
\usepackage{url}

\usepackage{siunitx}
\usepackage{subcaption} 
\usepackage{booktabs} 
\usepackage{mathtools}
\usepackage{multicol}
\usepackage{bm}
\usepackage{csquotes}
\usepackage{physics}
\usepackage{array}
\usepackage{multirow}
\usepackage{makecell}
\usepackage{cancel}
\usepackage{xcolor}
\colorlet{highlight}{cyan!50}
\usepackage{soul}
\sethlcolor{highlight}

\graphicspath{{./graphics/}}

\newcolumntype{L}[1]{>{\raggedright\let\newline\\\arraybackslash\hspace{0pt}}m{#1}}
\newcolumntype{C}[1]{>{\centering\let\newline\\\arraybackslash\hspace{0pt}}m{#1}}
\makeatletter
\renewcommand\section{\@startsection {section}{1}{\z@}%
	{18\p@ \@plus 6\p@ \@minus 3\p@}%
	{9\p@ \@plus 6\p@ \@minus 3\p@}%
	{\sffamily\normalsize\bfseries\boldmath}}
\makeatother

\begin{document}
\begin{frontmatter}

	\title{\sffamily\textbf{Inductive Position Sensors based on Coupling of Coils on Printed Circuit Boards for Demanding Automotive Applications}}
\author{\textbf{Stefan Kuntz\textsuperscript{1,2,*}\orcidlink{0000-0001-6364-6114}, Gerald Gerlach\textsuperscript{2}\orcidlink{0000-0002-7062-9598} and Sina Fella\textsuperscript{1}}\\[1\baselineskip]\textsuperscript{1}Robert Bosch GmbH, Abstatt, Germany
	\\\textsuperscript{2}Faculty of Electrical and Computer Engineering, Dresden University of Technology, Dresden, Germany
	}
\cortext[cor1]{\textit{Correspondence:} \texttt{stefan.kuntz@de.bosch.com}}
\abstracttitle{\sffamily Abstract}
\begin{abstract}
	Rotor position feedback is required in many industrial and automotive applications, e.g. for field-oriented control of brushless motors. Traditionally, magnetic sensors, resolvers or optical encoders are used to measure the rotor position. However, advances in inductive sensing concepts enable a low-cost, high-precision position measurement principle which is robust against magnetic stray fields exceeding 4000 A/m. The operating principle is based on the coupling of a transmitter coil with several receiver coils in the megahertz frequency range. The coils are part of a printed circuit board (PCB) which also comprises circuitry for demodulation and signal processing. The transmitter coil induces eddy currents in an electrically conductive passive coupling element, which provides position-dependent amplitude modulation. The voltage induced in the receiver coils encodes the rotor angle information, typically in quadrature signals. The coupling element requires no rare-earth materials and can be made of stainless steel, for instance. The PCB-based design of the sensor offers considerable flexibility in optimizing its performance. By tailoring the coil geometry and arrangement, accuracy, air gap and overall sensor dimensions can be adjusted to meet a broad range of application-specific requirements. A sensor design sample exhibits a mechanical angle error less than \ang{0.02} (\ang{0.1} electrical) in both, finite-element simulation and test bench measurement, with good agreement.
		\end{abstract}
\end{frontmatter}
    \section{Introduction}
\begin{figure}
    	\includegraphics[width=\columnwidth]{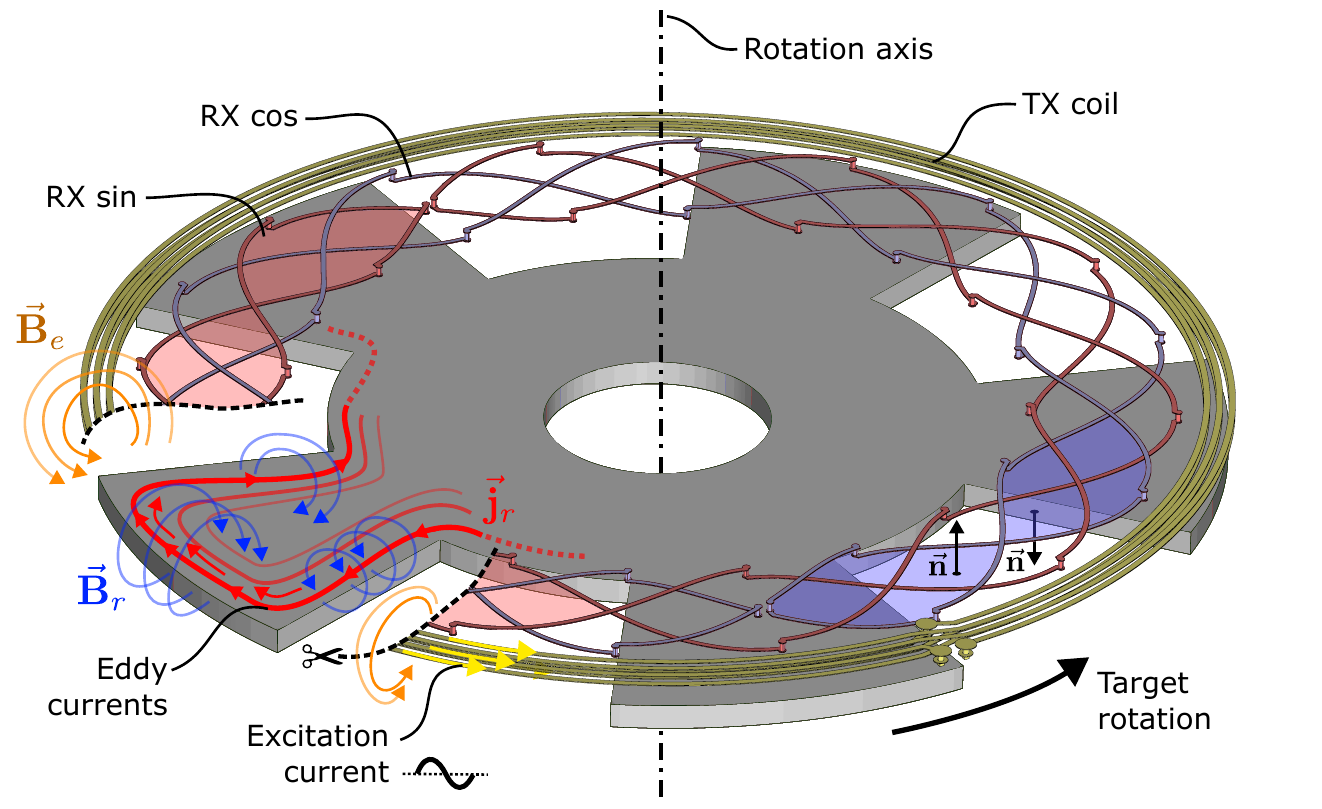}
    	\caption{Operating principle of a rotary inductive position sensor. An AC current in the megahertz frequency range flows through the transmitter coil (TX) and generates a magnetic excitation field $\mathbf{B}_e$ which induces eddy currents $\mathbf{j}_r$ in an electrically conductive target material, which in turn generates an opposing reaction field $\mathbf{B}_r$ that induces a voltage in the receiver coils (RX). Due to the shape of the target and the spatial phase shift between the two receiver coils, the target angle can be measured by calculating the angle of the demodulated receiver coil voltages. The differential nature of the RX coils suppresses direct coupling of TX coil into the RX coils due to the inverted surface normals $\mathbf{n}$ of the RX coil area segments. Figure published in \cite{Kuntz2024}.}
    	\label{fig:sensor-overview}
    \end{figure}

    The measurement of rotary position is an ancient problem in many disciplines of engineering, which was first solved by electromagnetic induction with a device called the \emph{Selsyn} or \emph{synchro} over one-hundred years ago \cite{Edward1921,Linville1934}. Initially used in electromechanical computer systems, over time the technology matured into \emph{resolvers} which can be seen as the direct predecessor of coupled-coil inductive position sensors. Specifically \emph{variable reluctance} (VR) resolvers---which are in use since at least the 1950s \cite{Kronacher1957}---share many similarities with inductive position sensors. The angle is encoded in amplitude-modulated signals in multiple receiver coils, while the excitation is provided by a separate coil. However, in inductive sensors the mechanism to vary the position-dependent coupling is based entirely on eddy-current effects (\autoref{fig:sensor-overview}) instead of a magnetic reluctance effect from ferromagnetic material, as in VR resolvers.
    
    This paper gives a brief introduction to inductive sensors and covers some basic design guidelines as well as an overview of how to assess the performance of a sensor designs. Initially introduced in the early 2000s \cite{Hobein2004}, inductive position sensors are highly relevant in automotive and industrial applications today. On the market, a trend can be observed for such sensors to replace more complex and potentially more expensive wound-wire resolvers, while on the other hand providing better performance than typical magnetic sensors which are commonly used as an alternative \cite{Datlinger2020}.
    
\pagestyle{plain}
\section{Design of an Inductive Position Sensor}
Fundamentally, an inductive angle position sensor consists of a transmitter coil (TX) together with an oscillator circuit for excitation (\autoref{fig:RLC}), at least two receiver coils (RX), an application-specific integrated circuit (ASIC, \autoref{fig:ASIC}) for signal processing, and a coupling element---the \emph{target} (\autoref{fig:sensor-overview}). All of the components have to be matched to ensure proper performance of the sensor.
\subsection{Overview}
Contrary to wound-wire resolvers, inductive position sensors are contained entirely on a printed circuit board (PCB) on the stator side. The PCB-based design enables adaptation to specific applications with small effort, while providing a high degree of integration. For example, directly embedding the sensor in an electric motors is possible. The simple and robust PCB-based design combined with intrinsic stray-field robustness are key features in replacing competing sensor technologies.
    
The coupling element on the rotor side---commonly referred to as the \emph{target}---simply consists of electrically conductive, preferably non-ferromagnetic material ($\mu_r\approx 1$), such as stainless steel, aluminum, or copper. Rare earth materials are therefore not required and are not beneficial to sensor operation. The target is typically thin in axial direction, rendering laminated sheet construction unnecessary.

The fundamental operating principle is based on eddy currents radial plane of the target, which are induced by the excitation coil at an operating frequency of typically \SI{3}{\mega\hertz} to \SI{5}{\mega\hertz}. The coupling between excitation and receiver coils varies with the target angle due to induced eddy currents, which provide a local \emph{damping} of the magnetic field of the TX coil. Due to the specifically shaped receiver coil area segments, and the spatial phase shift between the different receiver coils, the rotation angle of the target is encoded in the induced RX coil signals.

Due to the high excitation frequency in the megahertz range, the \emph{skin depth} $\delta$ \cite{Fawzi1985} in electrically conductive materials is rather small (\autoref{table:skin-depth}). In good approximation, the skin depth is determined by
\begin{equation}
	\delta = \sqrt{\frac{2}{\omega\sigma\mu}}
\end{equation}
where $\omega$ (\si{\per\second}) is the excitation frequency in radians per second, $\sigma$ the electrical conductivity (\si{\siemens\per\meter}), and $\mu$ the magnetic permeability (\si{\henry\per\meter}) of the material. The skin depth affects currents in the coils, as well as coupled eddy currents in the target material. Due to the skin effect, the eddy currents in the target primarily flow along the outline of the target (\autoref{fig:sensor-overview}). Ferromagnetic materials do not provide good performance as a target, since the smaller skin depth due to the higher magnetic permeability results in low signal amplitudes.

In contrast to \emph{incremental} angle encoders (e.g. optical encoders), inductive sensors are typically constructed to provide direct angle feedback without the need for an index pulse to resolve an angle ambiguity. In motor control applications, the periodicity $p$ of the sensor is often matched to the pole-pair count of the motor. The periodicity $p$ determines the number of receiver coil periods and the number of target wings. The angle obtained from the receiver coil signals can then be used for commutation immediately, and no initialization routine is required. A distinction therefore has to be made between the \emph{electrical angle} obtained directly from the sensor signals, and the mechanical angle---which differs from the mechanical angle by a factor of $p$. Within one full mechanical revolution, there are $p$ electrical revolutions of the sensor angle (\autoref{fig:signals-atan}). The range of uniqueness of the angle signal, given the periodicity~$p$, is therefore $2\pi/p$~radians.

Various other related variants of inductive position sensors exist, such as linear designs or C-shaped rotary sensors which only cover a part of the full circle in order to save space. Many of the same considerations apply to these sensors, however, their specific design is out of scope for the overview given here.

	\begin{table}
	\caption{Skin depth $\delta$ of common materials at relevant frequencies. Assuming $\sigma=\SI{35.38}{\mega\siemens\per\meter}$ for aluminum and $\sigma=\SI{1.4}{\mega\siemens\per\meter}$, $\mu_r=1$ for {1.4301} stainless steel. Even though the electrical conductivity $\sigma$ of ferromagnetic steel is much lower than $\sigma$ of aluminum or copper, the skin depth is much smaller due to the higher magnetic permeability. The relative permeability of steel generally decreases at higher frequencies \cite{Bowler2006}.}
	\label{table:skin-depth}
	\begin{tabular}{l r r r}
		\toprule
		Material & $f=\SI{3}{\mega\hertz}$ & $f=\SI{4}{\mega\hertz}$ & $f=\SI{5}{\mega\hertz}$\\\midrule
		Copper & \SI{38}{\micro\meter} & \SI{33}{\micro\meter}& \SI{30}{\micro\meter}\\
		Aluminum & \SI{49}{\micro\meter} & \SI{42}{\micro\meter}& \SI{38}{\micro\meter}\\
		1.4301 stainless & \SI{246}{\micro\meter} & \SI{213}{\micro\meter}& \SI{190}{\micro\meter}\\
		Ferromagn. steel & $\lesssim \SI{20}{\micro\meter}$ & $\lesssim\SI{20}{\micro\meter}$ & $\lesssim\SI{20}{\micro\meter}$\\\bottomrule
	\end{tabular}
\end{table}

\subsection{Printed Circuit Board (PCB)}
The PCB of the sensor contains coils in the form of copper traces. To avoid intersections, traces are arranged on several layers of the PCB, and are connected with \emph{vias} from one layer to another. The copper layers are separated by \emph{prepreg} dieletric.
For mechanical integrity, the PCB typically contains a thicker core in the middle, leading to a total thickness of typically \SI{0.8}{\mm} to \SI{1.6}{\mm}. In order to achieve best performance in terms of winding coupling and receiver coil voltage offset, the coils are typically located on one side of the PCB core to achieve small spacing between the coil copper layers. This side of the PCB then faces the target to minimize the air gap, i.e. the distance between the coil structures and target surface. PCBs with four layers are a therefore common choice for inductive position sensors, since the differential receiver coils structures require two layers to avoid intersections.

	\subsection{Transmitter Circuit}
\begin{figure}
		\centering
		\includegraphics[width=\columnwidth]{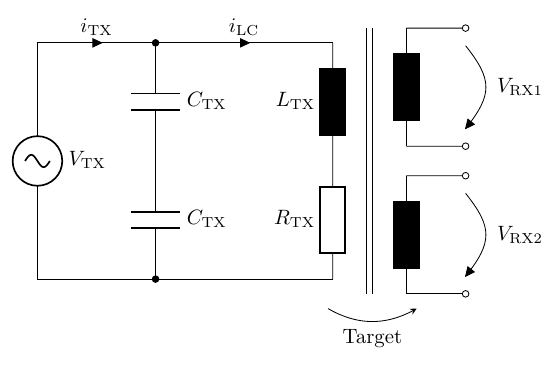}
		\caption{Simplified schematic of a coupled-coil inductive position sensor as an equivalent transformer. The transmitter (TX) coil on the PCB forms a parallel LC-tank together with two discrete capacitors $C_\mathrm{TX}$. The coil resistance $R_\mathrm{TX}$ is parasitic. The LC-oscillator is excited at the resonance frequency $\omega_0$. The \emph{target} (\autoref{fig:target}) acting as a coupling element provides a position-dependent mutual inductance between TX and RX coils, which leads to a varying output voltage amplitude of the modulated receiver coil signals $V_\mathrm{RX1},\,V_\mathrm{RX2}$.}
		\label{fig:RLC}
	\end{figure}

	In order to drive the excitation coil efficiently with a sufficiently large current for operation of the sensor at a frequency of approximately \SI{3}{\mega\hertz} to \SI{5}{\mega\hertz}, a parallel LC resonant circuit is constructed. The inductance is provided by the transmitter coil PCB trace structure itself, while discrete components are required for the capacitance. Ideally, the LC circuit (\autoref{fig:RLC}) is operated at its natural resonance frequency $\omega_0$.
	The resonant frequency of the parallel LC-tank with parasitic coil resistance $R_\mathrm{TX}$ (\autoref{fig:RLC}) is given by
	\begin{align}
		\label{eq:resonance}
		\omega_0 &= \sqrt{\frac{2}{L_\mathrm{TX}C_\mathrm{TX}} - \left(\frac{R_\mathrm{TX}}{L_\mathrm{TX}}\right)^2}
		\shortintertext{which simplifies to}
		\label{eq:resonance-simplified}
		\omega_0 &\approx \frac{\sqrt{2}}{\sqrt{L_\mathrm{TX}C_\mathrm{TX}}}
	\end{align}
	if the resistance $R$ is sufficiently small compared to the $LC$ contribution.
	The inductance is determined by the coil geometry, and the operating frequency range is constrained by the ASIC. Therefore, the capacitors $C_\mathrm{TX}$ can be most easily adjusted to match the constraints.
	\begin{equation}
		\label{eq:capacitance}
		C_\mathrm{TX} = \frac{2L_\mathrm{TX}}{L_\mathrm{TX}^2\omega_0^2 + R_\mathrm{TX}^2} \approx \frac{2}{L_\mathrm{TX}\omega_0^2}
	\end{equation}
	for a given targeted resonance frequency $\omega_0$.
	In practice, neglecting the resistance $R$ when estimating the resonance frequency is appropriate for transmitter coils with many tightly-coupled windings. For example, in order to obtain a resonance frequency of \SI{5}{\mega\hertz} given a typical TX inductance of \SI{3}{\micro\henry}, two \SI{680}{\pico\farad} capacitors in series (\autoref{fig:RLC}) would be an appropriate choice.
	
	Since current \emph{circulates} in the (R)LC-tank, only the resistive losses have to be resupplied by the transmitter coil driving circuit. The driving current $I_\mathrm{TX}$ is smaller than the total current $I_\mathrm{LC}$ flowing through the transmitter coil (\autoref{fig:RLC}). For AC constant-voltage excitation with $V_\mathrm{TX}$,
	\begin{equation}
		\label{eq:driving-current}
		I_\mathrm{TX} = \frac{V_\mathrm{TX}R_\mathrm{TX}C_\mathrm{TX}}{2L_\mathrm{TX}}\,,
	\end{equation}
	which can be derived by combining Eq.~\eqref{eq:resonance} together with the resonance condition of zero susceptibility.
	
	The complex total LC current is
	\begin{equation}
		i_\mathrm{LC} = \frac{V_\mathrm{TX}}{R_\mathrm{TX} + j\omega_0 L_\mathrm{TX}}\,.
	\end{equation}
	Since typically $\omega_0 L_\mathrm{TX} \gg R_\mathrm{TX}$, the imaginary part of $i_\mathrm{LC}$ dominates.
	In practice, the LC tank circuit has to be designed to limit $I_\mathrm{TX}$ according to the driving capability of the ASIC. Note that the above also applies to resistive losses in inductively coupled eddy current domains, such as the target. Due to the skin effect and proximity effect, the coil resistance at the operating frequency must be considered, which is larger than the DC resistance. Further, $R_\mathrm{TX}$ is strongly temperature dependent due to the temperature coefficient of the copper PCB trace resistivity, which is typically \SI{0.393}{\percent\per\kelvin}. Therefore, $R_\mathrm{TX}$ will be around $\SI{40}{\percent}$ larger at \SI{120}{\celsius} than at room temperature. Consequently, $R_\mathrm{TX}$ must be understood as an equivalent coil resistance, combining these effects.
	
		\begin{figure}
		\includegraphics[width=\columnwidth]{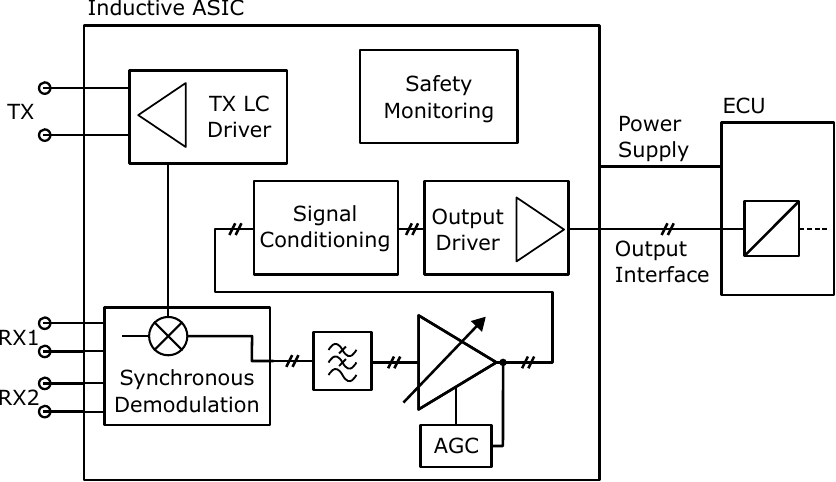}
		\caption{Simplified block diagram of an inductive ASIC. The TX driver provides an AC excitation current for the LC circuit with a resonant frequency in the range of \SI{3}{\mega\hertz} to \SI{5}{\mega\hertz}. The induced voltage in the receiver coils is demodulated synchronously with the TX excitation as reference. The receiver coil signal amplitude in the order of $\sim\SI{10}{\milli\volt}$ is amplified, typically with an automatic gain control loop. Typical outputs are analog differential sine/cosine signals or digital automotive interfaces such as SENT.}
		\label{fig:ASIC}
	\end{figure}
	
	Typically, circular transmitter coils with several windings in series are used, which surround the receiver coils and target. The main parameter that must be chosen for the TX coil is the number of windings. Assuming perfect coupling of a number of $N$ transmitter coil windings---which is an appropriate assumption since the PCB layer distance, PCB trace width, and trace distance are typically small---the inductance grows with the squared number of windings, i.e. $L_\mathrm{TX} \approx \mathcal{O}\left(N^2\right)$. The resistance $R_\mathrm{TX}$ is mainly dependent on the length of the overall coil trace, therefore $R_\mathrm{TX} \approx \mathcal{O}\left(N\right)$ when neglecting winding proximity effects. Consequently, $I_\mathrm{TX} \approx \mathcal{O}\left(\frac{1}{N}\right)$. Further, since $\omega_0 L_\mathrm{TX} \gg R_\mathrm{TX}$ due to large $\omega_0$, the current in the LC circuit is mainly limited by the inductance. Assuming constant-voltage AC excitation of the LC circuit, $I_\mathrm{LC} \approx \mathcal{O}\left(\frac{1}{N^2}\right)$. However, since $I_\mathrm{LC}$ flows in $N$ windings, the effective magnetic excitation field for the target is $\mathcal{O}\left(\frac{1}{N}\right)$. Assuming linearity, the same relation with regard to the dependence on $N$ then holds for the induced receiver coil voltages at constant excitation frequency.
	
	Therefore, the choice of the number of TX windings $N$ presents a trade-off: Choose $N$ large enough so that $I_\mathrm{TX}$ is acceptably below the driver limit, but small enough so that the signal amplitude is sufficient.
	
	\begin{figure}
		\centering
		\includegraphics[width=\columnwidth]{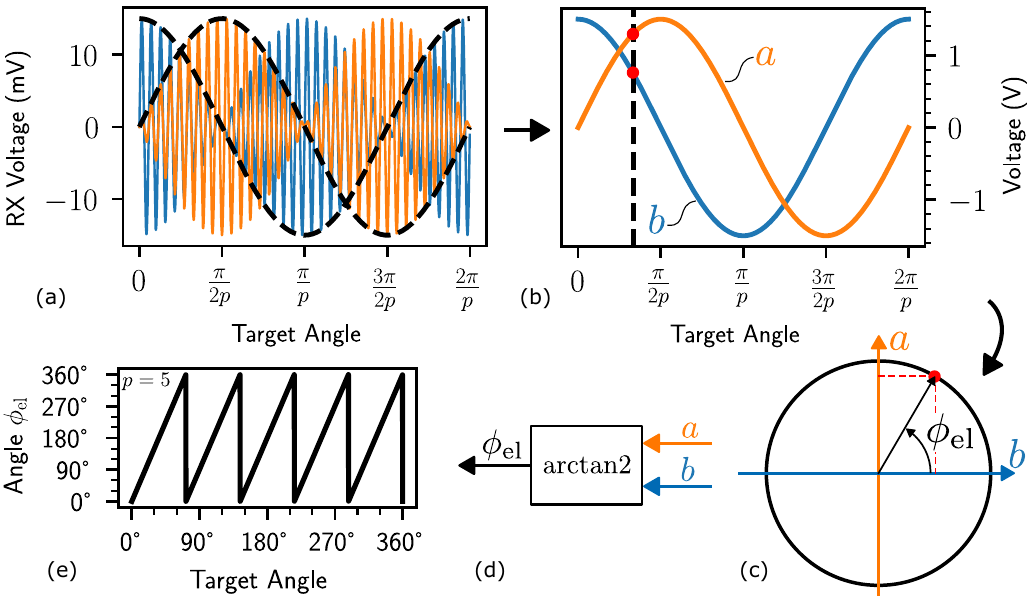}
		\caption{Signal processing of inductive position sensor signals. The induced voltage (a) in the receiver coils is demodulated synchronously with the TX voltage (\autoref{fig:RLC}) and amplified (b). In the Lissajous figure, the signals $a$ and $b$ approximately trace a circle (c). The encoded angle can be recovered with the arctangent function (d). The resulting electrical angle repeats $p$-times within a full mechanical rotation (e), according to the periodicity $p$ of the sensor.
		}
		\label{fig:signals-atan}
	\end{figure}
	
	\subsection{Receiver Coils and Target}
	The receiver coil design is an essential part of the development of an inductive position sensor, since the RX coils are the most important factor in determining the intrinsic angle error performance of the sensor, together with the target. Since the topic is complex, we refer to existing literature for an in-depth discussion of receiver coil design \cite{Kuntz2024,Dauth2023a}.
	
	\begin{figure}
		\centering
		\begin{subfigure}[b]{0.48\columnwidth}
			\centering
			\includegraphics[width=\textwidth]{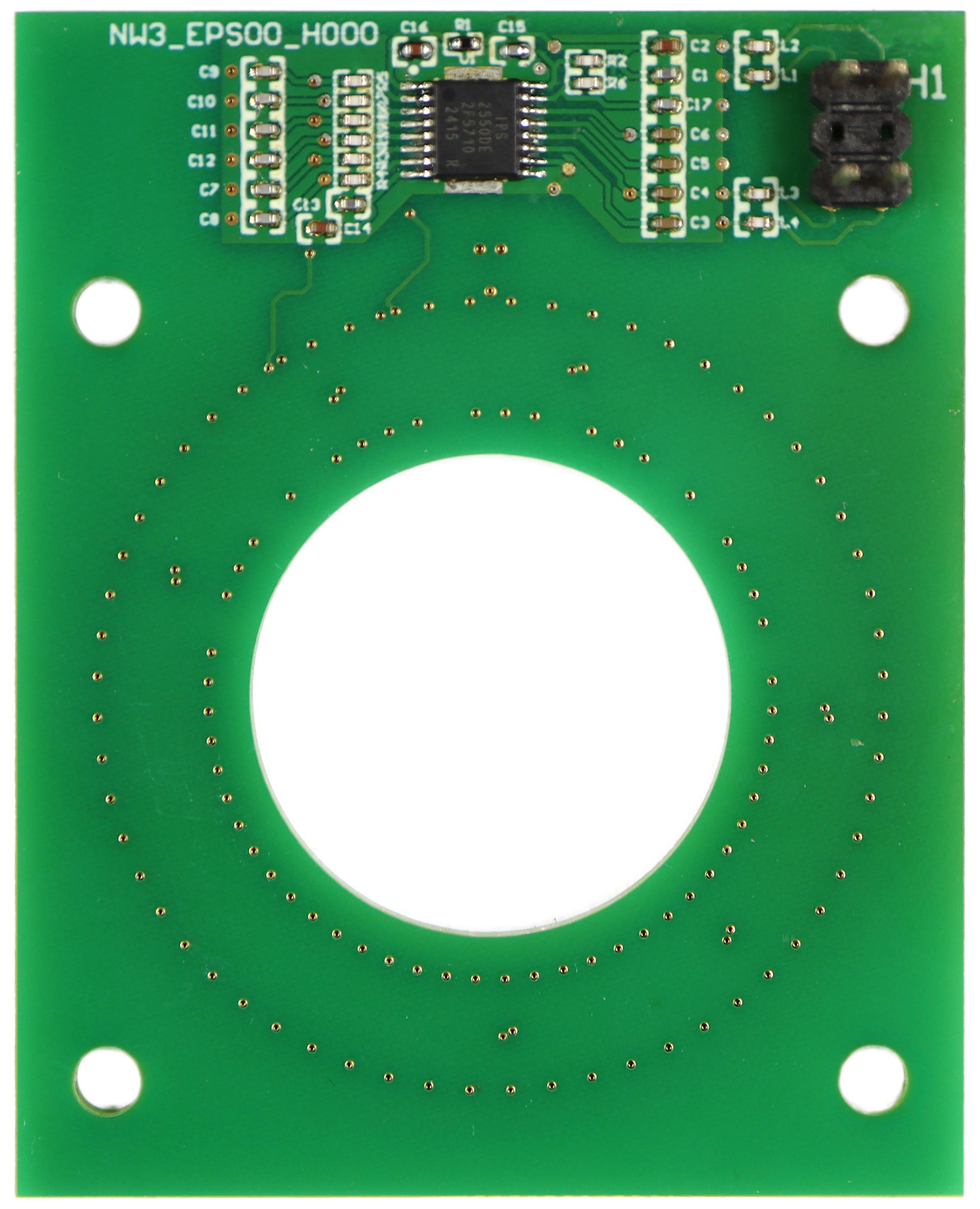}
			\caption{}
		\end{subfigure}\hfill
		\begin{subfigure}[b]{0.48\columnwidth}
			\centering
			\includegraphics[width=\textwidth]{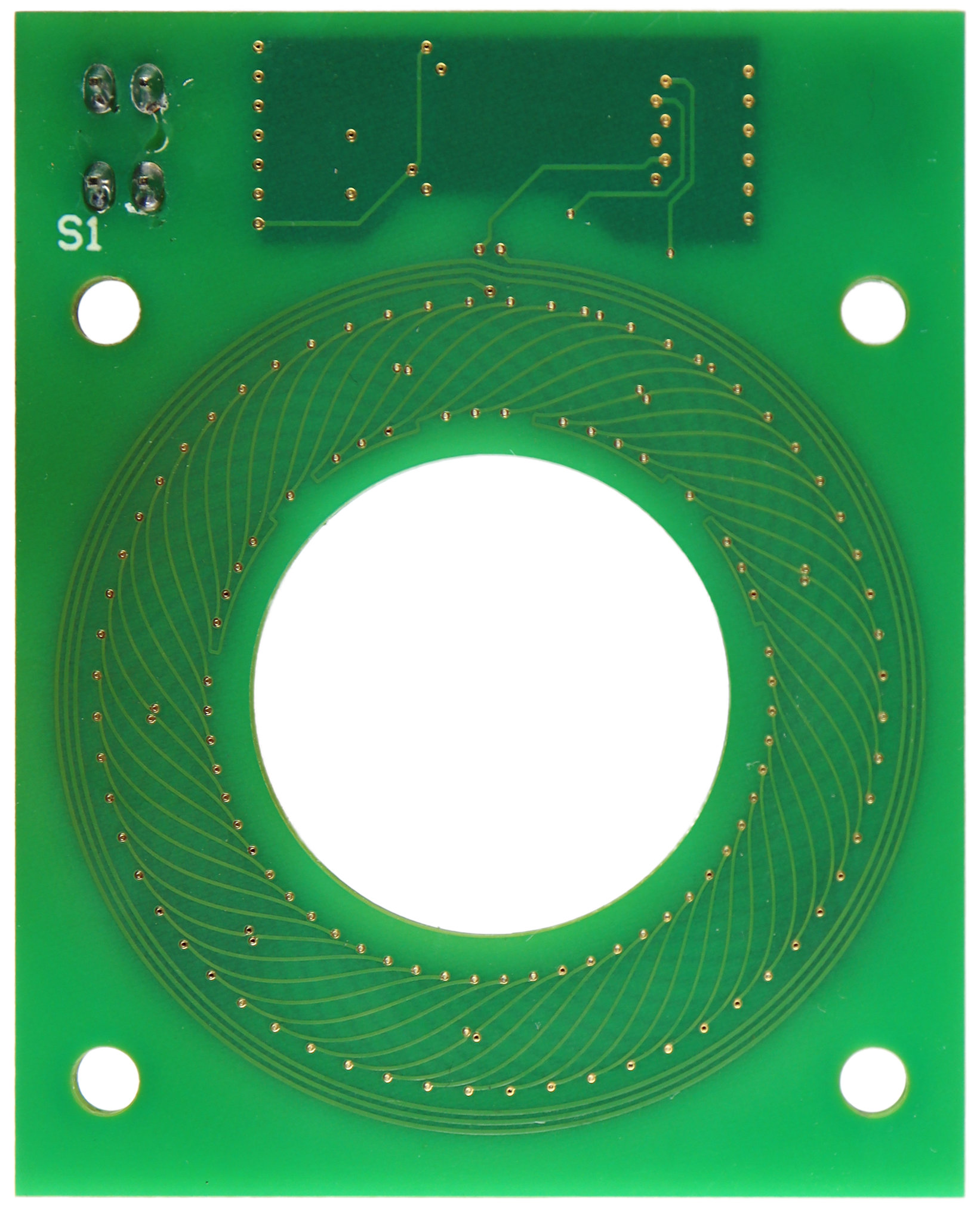}
			\caption{}
		\end{subfigure}
		\caption{Inductive sensor printed circuit board (PCB) sample for research purposes. Top side (a) with ASIC, passive components, and connector; bottom side (b) with coil system, facing the target. The PCB consists of four layers, the bottom two layers are used for the coil system. The two receiver coils have periodicity $p=5$ and three windings each ($n_w=3$) with a cosine shape function. The hole in the center of the PCB allows for end-of-shaft as well as on-axis applications. This sensor provides single-ended non-differential analog sine/cosine output signals on a 4-pin interface together with \SI{5}{\volt} supply and ground pins.}
		\label{fig:PCBs}
	\end{figure}

	The primary concern in many applications is reaching a sufficient amplitude of the induced receiver coil voltage, especially if the available space for the sensor is constrained. A smaller sensor size and therefore smaller receiver coil surface area will lead to a smaller induced voltage. An increased air gap between coil structure and target surface leads to the same effect. Smaller signal amplitudes require higher amplification factors in the ASIC, resulting in a worse signal-to-noise ratio.
	
	RX coils are \emph{differential} by alternating the surface normal of the area enclosed by the receiver coil (shaded area in \autoref{fig:sensor-overview}). Ideally, there is negligible direct coupling from the TX coil to the receiver coils in the absence of a target. The specific shape of the receiver coil area segments combined with the target wings introduce an imbalance in the otherwise differential coil structure, which leads to an induced voltage. This effect can be understood as a local damping due to the magnetic field of the eddy currents flowing in the target (\autoref{fig:sensor-overview}). To obtain orthogonal signals for angle calculation, the geometry of a receiver coil is duplicated and spatially phase-shifted. This also phase-shifts the demodulated induced voltage w.r.t. the target angle. In two-phase designs, the phase shift is obtained by rotating the second receiver coil by $\ang{90}/p$ relative to the first coil.
	
	The shape of the receiver coil has a significant effect on the angle error. Minimal error is generally achieved with a sinusoidal shape function for the traces of the receiver coils. In this case, a cosine function is used to generate the centerline of the trace. The design process is explained in detail in \cite{Kuntz2024}, including the relevant coordinate transforms, via placement and centerline function parametrization. As a rule of thumb, the area spanned by the receiver coil centerline function should cover the target wings, but not exceed them in radial direction.
	
\begin{figure}
		\centering
		\begin{subfigure}[b]{0.48\columnwidth}
			\centering
			\includegraphics[width=\textwidth]{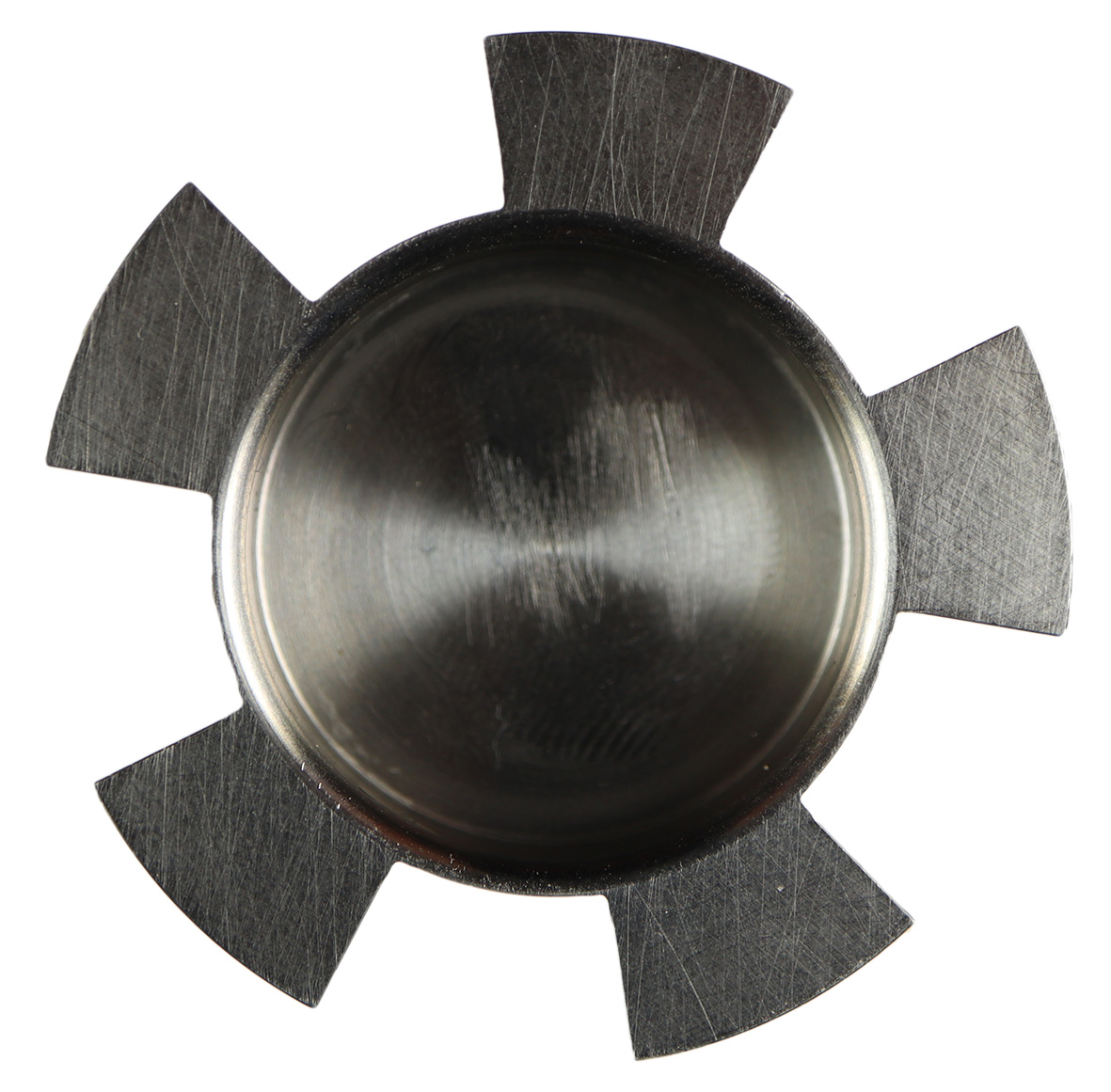}
			\caption{}
		\end{subfigure}\hfill
		\begin{subfigure}[b]{0.48\columnwidth}
			\centering
			\includegraphics[width=\textwidth]{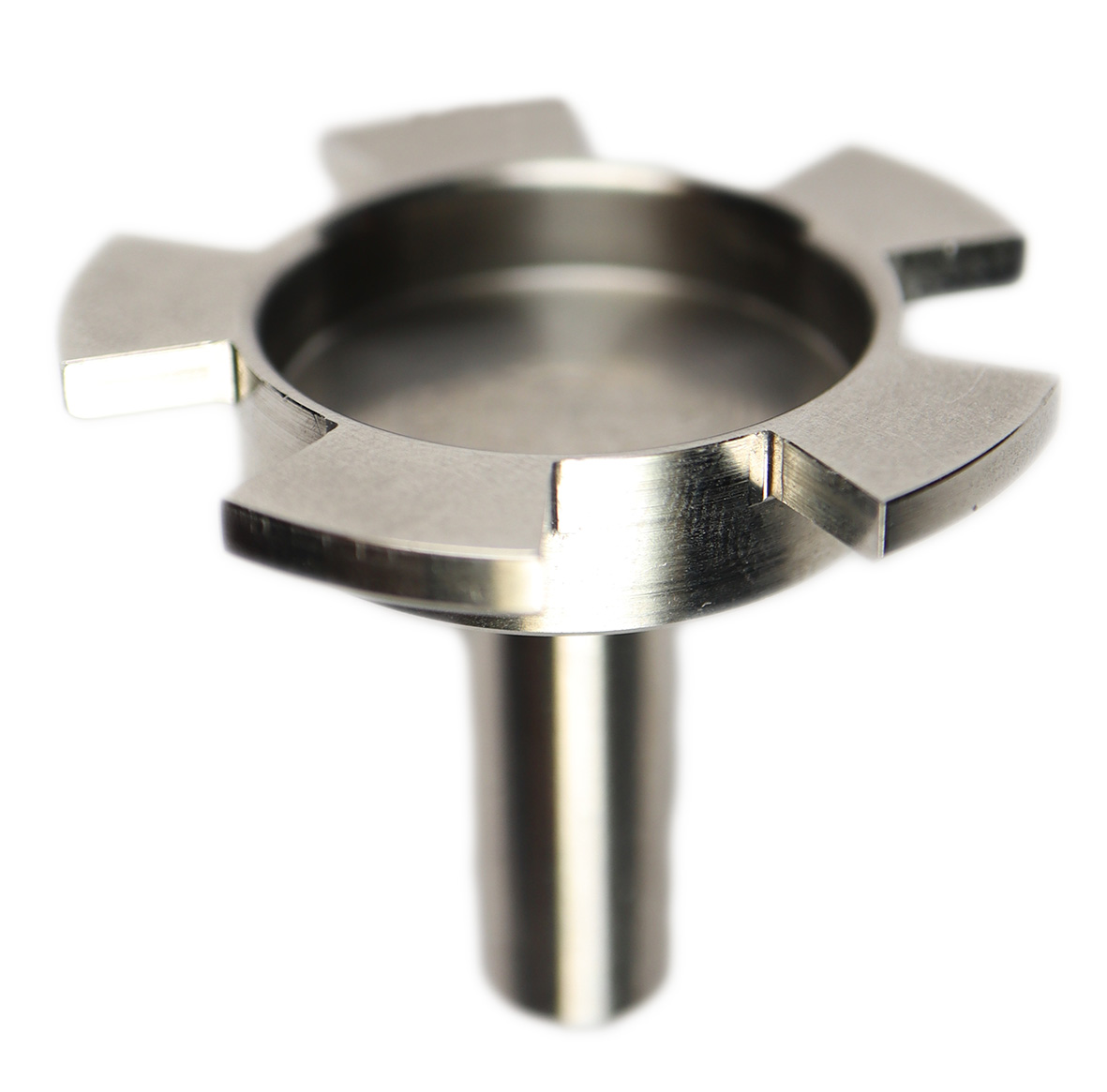}
			\caption{}
		\end{subfigure}
		\caption{Stainless steel rotary target milled from a solid block, used for testing purposes with the PCB from \autoref{fig:PCBs}. (a) view from the sensor PCB side, (b) side view with shaft. The target wing diameter is \SI{45}{\mm}/\SI{27}{\mm}, the wing thickness is \SI{3}{\mm}. In practice, stamped and bent sheet metals parts are often used instead. The number of \emph{wings} matches the periodicity $p=5$ of the sensor.}
		\label{fig:target}
	\end{figure}

	The example sensor design depicted in Figures~\ref{fig:PCBs} and \ref{fig:target} provides a voltage transfer ratio between transmitter and receiver coil of approximately $\SI{1}{\volt}/\SI{20}{\volt}$ at an air gap of \SI{1}{\mm}. The rather large transfer ratio is achieved due to the use of three receiver coil windings in series ($n_w=3$). Therefore, the sensor is also able to operate at much larger air gap distance between PCB and target. Typical inductive ASICs are able to operate down to approximately $\SI{1}{\volt}/\SI{250}{\volt}$ without significant loss of accuracy.
	
	The use of multiple phase-shifted RX windings in series has two significant benefits: The amplitude of the induced voltage is \emph{increased}, while the angle error is \emph{decreased} by destructive interference of unwanted higher-order harmonics in the receiver coil signals \cite{Kuntz2024}.
	By connecting the windings with differential structures, the impact of the series connection structure on the signals is negligible. The existing PCB layers are reused and shared between coils by placing additional windings in-between existing windings with an appropriate phase shift. Since this is done for all receiver coils, the phase of the resulting RX signals relative to each other is not affected. The number of windings can be increased until the design reaches the limits of the PCB manufacturing process, e.g. the minimum distance between the annular ring of vias and coil traces. Additionally, RX coils can also be duplicated on additional PCB layers and connected in series, additionally boosting the amplitude.
	
	\subsection{Signal Processing}
	Although it would be possible to construct a signal processing circuitry for an inductive position sensor from discrete parts, real-world application use an application-specific integrated circuit (ASIC) that combines the required signal processing blocks into a single package (\autoref{fig:ASIC}). 	In the recent years, the number of ASICs available on the market
	\cite[e.g.,][]{AMS2024,Melexis2022,SCI2022,REC2022,TexasInstruments2023} increased considerably. While some of their features and implementation details differ, most inductive ICs include the same key components (\autoref{fig:ASIC}).
	
	An inductive position sensor ASIC comprises a TX driver, which is able to drive the LC oscillator circuit at its resonance frequency and supplies the $I_\mathrm{TX}$ current. The voltages induced in the receiver coils are demodulated synchronously with the TX excitation phase as the reference \cite{Rahal2009b}. Since the receiver coil signals are generally quite weak, in the range of approximately \SI{5}{\milli\volt} to \SI{100}{\milli\volt} amplitude, an amplification stage is employed to boost the signal amplitudes. To compensate variations of the air gap, an automatic gain control (AGC) loop is often used.
	
	Depending on the ASIC, further signal conditioning may take place in a digital part of the IC after digitalization with an ADC. In this case, on-chip angle calculation is possible (\autoref{fig:signals-atan}). Other ASIC implementations work entirely in the analog domain, requiring off-chip angle calculation in the ECU. Some ASICs work with three-phase coil systems which requires a Clarke transform to obtain orthogonal signals before angle calculation.
	
	Available ASICs on the market are predominantly ready for ASIL-B or ASIL-C functional safety applications as a Safety Element out of context (SEooC) \cite{ISO26262}. Safety features, such as detection of a broken PCB trace or short circuits, are typically included directly in the ASIC, which signals the safe-state to the electronic control unit (ECU).
	
	The output interface of the ASIC depends on the targeted application. For high-speed rotor position measurement for motor control, differential analog output signals are still common. For applications where a lower update rate is acceptable, the digital SENT interface \cite{J27162016} is a common choice in automotive environments.
	
	The inductive sensing principle exhibits a high degree of intrinsic stray-field robustness \cite{Brajon2022}. This is mainly due to two primary reasons: synchronous demodulation of the receiver coil signals with small bandwidth, and RX coil design. For example, in order to accommodate a maximum rotation speed of \num{300000} electrical rpm, a bandwidth of approximately $\pm\SI{5}{\kilo\hertz}$ around the excitation frequency is required. Secondly, the receiver coil geometry is inherently \emph{differential}. Therefore homogeneous stray fields are predominantly rejected, even if the frequency of the stray field matches the operating frequency of the sensor.

	\subsection{Finite Element Simulation}
	\begin{figure}
		\includegraphics[width=\columnwidth]{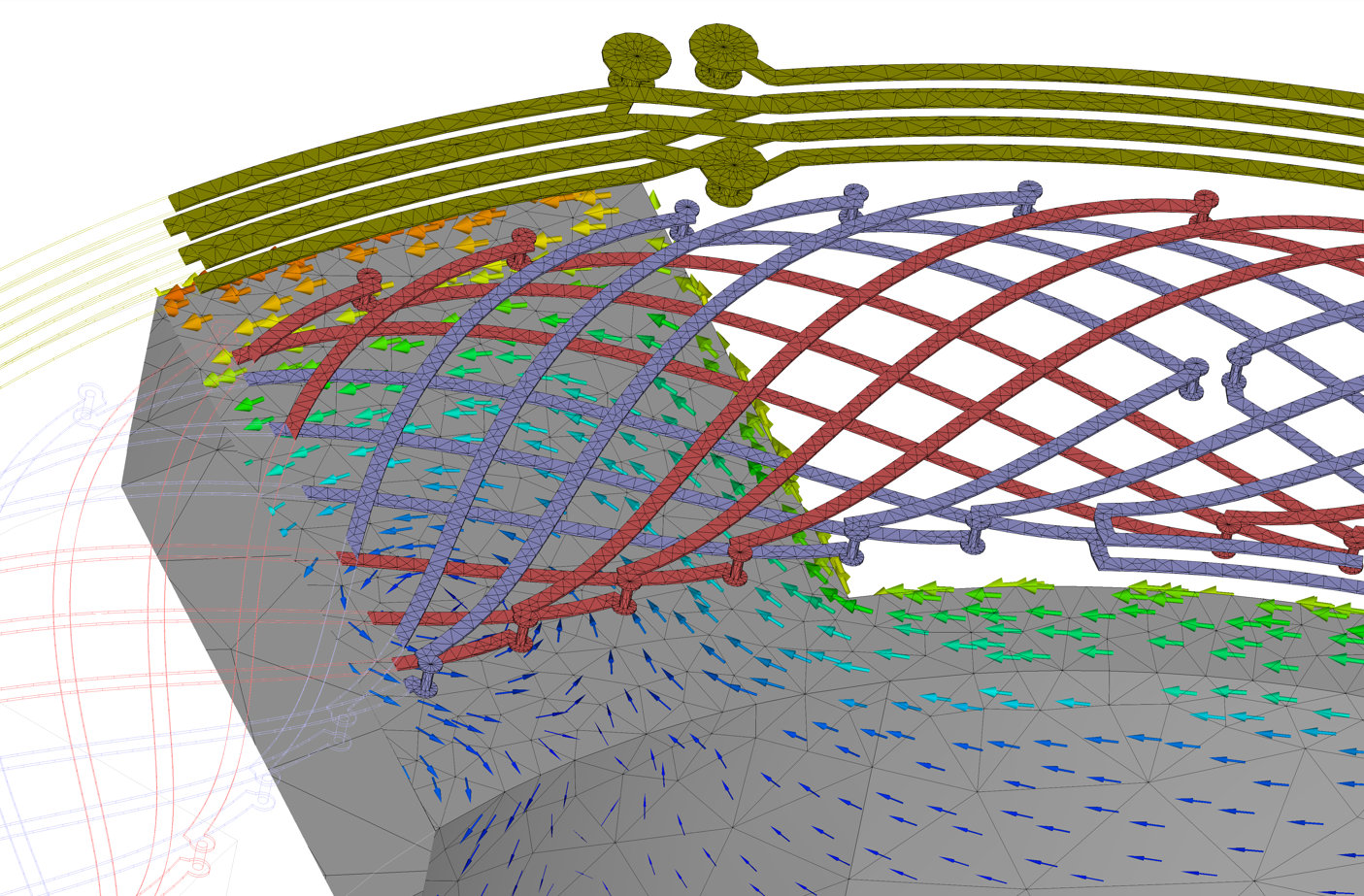}
		\caption{Finite element simulation of sensor geometry by solving a time-harmonic eddy current problem. The target is modeled as a surface mesh with impedance boundary condition, resulting in a hollow target shell. Coils and air are part of the volume mesh. The vector plot shows the surface current density of the target geometry due to the excitation current in the transmitter coil.}
		\label{fig:FEM}
	\end{figure}
	In order to assess the feasibility and performance of an inductive sensor design early in the development process, simulations with the finite element method (FEM) are commonly performed.	
	A number of useful approximations and simplifications can be made, known as the eddy current model of Maxwell's equations. The wavelength at \SI{4}{\mega\hertz} in vacuum is \SI{75}{\meter}, while the typical sensor size is in the order of \SI{100}{\milli\meter}. Therefore, radiation effects can be neglected. Since no ferromagnetic materials are used, the posed eddy current problem is linear. Due to excitation only at the resonant frequency the problem can be treated as time-harmonic. The target is typically a good conductor with small skin depth, therefore an impedance boundary condition on the target surface (\autoref{fig:FEM}) can be used in the eddy current model \cite{Fawzi1985}. Otherwise, a large number of volume elements would be required to accurately resolve the skin depth.
	
\begin{figure}
		\includegraphics[width=\columnwidth]{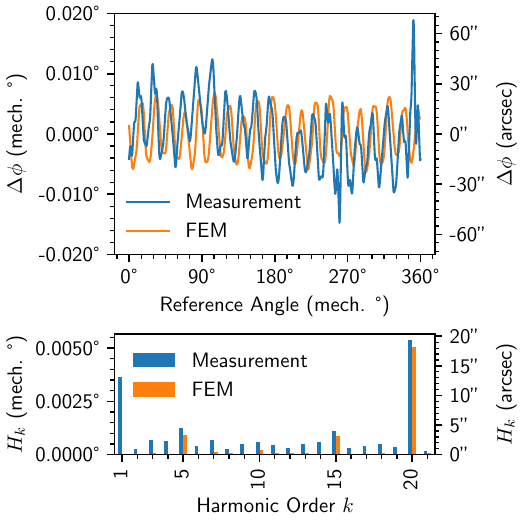}
		\caption{Angle error $\Delta\phi$ and harmonic angle error amplitudes $H_k$ of the sensor example from Figures~\ref{fig:PCBs} and \ref{fig:target} at \SI{1}{\mm} air gap. The finite-element simulation result from \cite{Kuntz2024} is given for reference---note that target dimensions of \cite{Kuntz2024} and the measurement differ in inner diameter and wing thickness (\autoref{fig:target}), though the coil system is identical.}
		\label{fig:nw3-error}
	\end{figure}

	The TX coil geometry is typically modeled as a solid, voltage-driven coil (\autoref{fig:RLC}), while the current in the receiver coils can be set to zero since the input impedance of the ASIC is usually sufficiently large. It is important to include the target itself and all metallic surrounding materials, such as the sensor housing, etc., in the simulation. Electrically conductive surrounding materials may have a significant effect on the RX amplitude and inductance of the transmitter coil, depending on the distance to the coil system. A typical mesh to resolve even very small angle errors (as in \autoref{fig:nw3-error}) typically consists around one million elements, with a strong dependence on the number of windings of the receiver coil.
	
	In motor control applications, the harmonic decomposition of the angle error is an important parameter to ensure good performance of the system. Inductive rotary position sensors with two receiver coils typically exhibit a dominant 4th electrical harmonic order. Since the RX coils are essentially phase-shifted copies, the phase relations of 3rd and 5th order distortions of the sine/cosine signals both lead to a 4th harmonic in the angle error signal, which can be derived using the framework established in \cite{Kuntz2022}.

	FEM simulation and measurement of the sample PCB (\autoref{fig:PCBs}) are in good agreement with regard to the observed angle error (\autoref{fig:nw3-error}). Both measurement and FEM signals are compensated for offset, amplitude mismatch and orthogonality before error calculation---as commonly done in ECU signal processing. The resuling angle error is below \ang{0.01} mechanically for the FEM simulation and below \ang{0.02} mechanically for the measurement result, considering the factor $p=5$ between mechanical and electrical angles. The peak-to-peak error in the measurement is larger due some signal disturbance, though the amplitude of the 4th electrical harmonic $H_{20}$ (20th mechanical order, since $p=5$) of the angle error matches much more closely between FEM and measurement, with an amplitude of approximately 20 arcseconds mechanically. Note that in this sample, no specific measures were taken to minimize the 4th harmonic contribution. The 4th harmonic can be suppressed further if desired, either by optimization of the sensor or by harmonic compensation in the ECU. The small first harmonic $H_1$ is caused by mechanical imperfections in the sample measurement setup which are not present in the ideal FEM simulation.
	
	Generally, electromagnetic finite element simulation is a suitable tool to support or even replace sample analysis for inductive position sensor development.

	\section{Summary}
	Inductive position sensors based on coupled-coils are a viable alternative to traditional position sensing concepts in a variety of applications, such as rotary position sensors for motor control. The PCB-based design concept is based on proven technology and offers excellent performance. Since the dimensions of the coil system and target can be freely chosen within sensible limits, the design approach is flexible and able to precisely fit application requirements. The sensor can be sized to exactly fit the available installation space. Further, advances in PCB manufacturing technology, such as tiny lasered microvias, allow for very small sensor sizes. Future research may benefit from an in-depth analysis of tolerance scenarios to assess the robustness of sensor designs with regard to mechanical misalignment and manufacturing tolerances.
	\vspace{-5pt}
	\section*{Acknowledgments}
	We thank Arne Fränznick for creating the ASIC PCB layout of the sample, and for conducting the measurements.
	\vspace{-5pt}
	\bibliography{./symposium-literature}

\begin{thebibliography}{19}
\expandafter\ifx\csname natexlab\endcsname\relax\def\natexlab#1{#1}\fi
\providecommand{\url}[1]{\texttt{#1}}
\providecommand{\href}[2]{#2}
\providecommand{\path}[1]{#1}
\providecommand{\DOIprefix}{doi:}
\providecommand{\ArXivprefix}{arXiv:}
\providecommand{\URLprefix}{URL: }
\providecommand{\Pubmedprefix}{pmid:}
\providecommand{\doi}[1]{\href{http://dx.doi.org/#1}{\path{#1}}}
\providecommand{\Pubmed}[1]{\href{pmid:#1}{\path{#1}}}
\providecommand{\bibinfo}[2]{#2}
\ifx\xfnm\relax \def\xfnm[#1]{\unskip,\space#1}\fi
\bibitem[{Edward and Waldo(1921)}]{Edward1921}
\bibinfo{author}{M.~H. Edward}, \bibinfo{author}{W.~W. Waldo},
  \bibinfo{title}{System for the transmission of angular movements},
  \bibinfo{year}{1921}. \bibinfo{note}{{Patent US1612117}}.
\bibitem[{Linville and Woodward(1934)}]{Linville1934}
\bibinfo{author}{T.~M. Linville}, \bibinfo{author}{J.~S. Woodward},
\newblock \bibinfo{title}{Selsyn instruments for position systems},
\newblock \bibinfo{journal}{Transactions of the American Institute of
  Electrical Engineers} \bibinfo{volume}{53} (\bibinfo{year}{1934})
  \bibinfo{pages}{953--960}.
\bibitem[{Kronacher(1957)}]{Kronacher1957}
\bibinfo{author}{G.~Kronacher},
\newblock \bibinfo{title}{Design, performance and application of the vernier
  resolver},
\newblock \bibinfo{journal}{Bell System Technical Journal} \bibinfo{volume}{36}
  (\bibinfo{year}{1957}) \bibinfo{pages}{1487--1500}.
\bibitem[{Kuntz et~al.(2024)Kuntz, Gerber, Gerlach, and Fella}]{Kuntz2024}
\bibinfo{author}{S.~Kuntz}, \bibinfo{author}{D.~Gerber},
  \bibinfo{author}{G.~Gerlach}, \bibinfo{author}{S.~Fella},
\newblock \bibinfo{title}{Design and analysis of receiver coils with multiple
  in-series windings for inductive eddy current angle position sensors based on
  coupling of coils on printed circuit boards},
\newblock \bibinfo{journal}{Sensors} \bibinfo{volume}{24}
  (\bibinfo{year}{2024}) \bibinfo{pages}{4880}.
\bibitem[{Hobein et~al.(2004)Hobein, Dori{\ss}en, and D{\"u}rkopp}]{Hobein2004}
\bibinfo{author}{D.~Hobein}, \bibinfo{author}{T.~Dori{\ss}en},
  \bibinfo{author}{K.~D{\"u}rkopp}, \bibinfo{title}{Progress in Automotive
  Position Sensors and Introduction of the Hella Inductive Position Sensor},
  \bibinfo{type}{Technical Report}, \bibinfo{year}{2004}. \bibinfo{note}{{SAE
  Technical Paper 2004-01-1115}}.
\bibitem[{Datlinger and Hirz(2020)}]{Datlinger2020}
\bibinfo{author}{C.~Datlinger}, \bibinfo{author}{M.~Hirz},
\newblock \bibinfo{title}{Benchmark of rotor position sensor technologies for
  application in automotive electric drive trains},
\newblock \bibinfo{journal}{Electronics} \bibinfo{volume}{9}
  (\bibinfo{year}{2020}) \bibinfo{pages}{1063}.
\bibitem[{Fawzi et~al.(1985)Fawzi, Ahmed, and Burke}]{Fawzi1985}
\bibinfo{author}{T.~Fawzi}, \bibinfo{author}{M.~R. Ahmed},
  \bibinfo{author}{P.~Burke},
\newblock \bibinfo{title}{On the use of the impedance boundary conditions in
  eddy current problems},
\newblock \bibinfo{journal}{IEEE Transactions on Magnetics}
  \bibinfo{volume}{21} (\bibinfo{year}{1985}) \bibinfo{pages}{1835--1840}.
\bibitem[{Bowler(2006)}]{Bowler2006}
\bibinfo{author}{N.~Bowler},
\newblock \bibinfo{title}{Frequency-dependence of relative permeability in
  steel},
\newblock in: \bibinfo{booktitle}{AIP Conference Proceedings}, volume
  \bibinfo{volume}{820}, \bibinfo{publisher}{AIP}, \bibinfo{year}{2006}, pp.
  \bibinfo{pages}{1269--1276}.
\bibitem[{Dauth et~al.(2023)Dauth, Gerlach, and Fella}]{Dauth2023a}
\bibinfo{author}{R.~Dauth}, \bibinfo{author}{G.~Gerlach},
  \bibinfo{author}{S.~Fella},
\newblock \bibinfo{title}{Inductive coupled-coils angle encoders with improved
  performance and linearity: Induktive winkelsensoren auf basis gekoppelter
  spulen mit erhöhter performanz und linearität},
\newblock \bibinfo{journal}{tm - Technisches Messen} \bibinfo{volume}{90}
  (\bibinfo{year}{2023}) \bibinfo{pages}{2--7}.
\bibitem[{{ams AG}(2024)}]{AMS2024}
\bibinfo{author}{{ams AG}}, \bibinfo{title}{{AS5715A/AS5715R On-/Off-Axis
  Inductive Position Sensor with Sin/Cos Output}}, \bibinfo{year}{2024}.
  \URLprefix \url{https://look.ams-osram.com/m/3f1bb9952af75d6b/},
  \bibinfo{note}{{Datasheet Rev. v5-00}}.
\bibitem[{{Melexis}(2022)}]{Melexis2022}
\bibinfo{author}{{Melexis}}, \bibinfo{title}{{MLX90510} high-speed inductive
  position sensor}, \bibinfo{year}{2022}. \URLprefix
  \url{https://media.melexis.com/-/media/files/documents/datasheets/mlx90510-datasheet-melexis.pdf},
  \bibinfo{note}{{Datasheet Rev. 003, 3901090510}}.
\bibitem[{{Semiconductor Components Industries}(2022)}]{SCI2022}
\bibinfo{author}{{Semiconductor Components Industries}},
  \bibinfo{title}{Inductive position sensor interface {NCV77320}},
  \bibinfo{year}{2022}. \URLprefix
  \url{https://www.onsemi.com/download/data-sheet/pdf/ncv77320-d.pdf},
  \bibinfo{note}{{Datasheet Rev. 2}}.
\bibitem[{{Renesas Electronics Corporation}(2024)}]{REC2022}
\bibinfo{author}{{Renesas Electronics Corporation}}, \bibinfo{title}{{Inductive
  Position Sensor IC IPS2550 Datasheet}}, \bibinfo{year}{2024}. \URLprefix
  \url{https://www.renesas.com/us/en/document/dst/ips2550-datasheet},
  \bibinfo{note}{{Datasheet Rev. April 24, 2024}}.
\bibitem[{{Texas Instruments}(2023)}]{TexasInstruments2023}
\bibinfo{author}{{Texas Instruments}}, \bibinfo{title}{{LDC5072-Q1} inductive
  position sensor front-end with sin/cos interface}, \bibinfo{year}{2023}.
  \URLprefix \url{https://www.ti.com/lit/ds/symlink/ldc5072-q1.pdf},
  \bibinfo{note}{{Datasheet Rev. SNOSD47C}}.
\bibitem[{Rahal and Demosthenous(2009)}]{Rahal2009b}
\bibinfo{author}{M.~Rahal}, \bibinfo{author}{A.~Demosthenous},
\newblock \bibinfo{title}{An integrated signal conditioner for high-frequency
  inductive position sensors},
\newblock \bibinfo{journal}{Measurement Science and Technology}
  \bibinfo{volume}{21} (\bibinfo{year}{2009}) \bibinfo{pages}{015203}.
\bibitem[{{International Organization for Standardization}(2018)}]{ISO26262}
\bibinfo{author}{{International Organization for Standardization}},
  \bibinfo{title}{{Road vehicles — Functional safety}}, \bibinfo{year}{2018}.
  \bibinfo{note}{{ISO Standard 26262-1:2018}}.
\bibitem[{{SAE International}(2016)}]{J27162016}
\bibinfo{author}{{SAE International}}, \bibinfo{title}{{SENT - Single Edge
  Nibble Transmission for Automotive Applications}}, \bibinfo{year}{2016}.
  \bibinfo{note}{{SAE Standard J2716\_201604}}.
\bibitem[{Brajon et~al.(2022)Brajon, Lugani, and Close}]{Brajon2022}
\bibinfo{author}{B.~Brajon}, \bibinfo{author}{L.~Lugani},
  \bibinfo{author}{G.~Close},
\newblock \bibinfo{title}{Hybrid magnetic–inductive angular sensor with 360°
  range and stray-field immunity},
\newblock \bibinfo{journal}{Sensors} \bibinfo{volume}{22}
  (\bibinfo{year}{2022}) \bibinfo{pages}{2153}.
\bibitem[{Kuntz et~al.(2022)Kuntz, Dauth, Gerlach, Ott, and Fella}]{Kuntz2022}
\bibinfo{author}{S.~Kuntz}, \bibinfo{author}{R.~Dauth},
  \bibinfo{author}{G.~Gerlach}, \bibinfo{author}{P.~Ott},
  \bibinfo{author}{S.~Fella},
\newblock \bibinfo{title}{Harmonic analysis of the arctangent function
  regarding the angular error introduced by superimposed {Fourier} series for
  application in sine/cosine angle encoders},
\newblock \bibinfo{journal}{Sensors and Actuators A: Physical}
  \bibinfo{volume}{344} (\bibinfo{year}{2022}) \bibinfo{pages}{113585}.

\end{thebibliography}
\end{document}